\documentclass[12pt]{article}
\usepackage[utf8]{inputenc}
\usepackage{amsmath,amsfonts,amssymb}
\usepackage{graphicx}
\usepackage{geometry}
\usepackage{authblk}
\usepackage{float}
 \usepackage{xcolor}

{\sffamily \title{\textbf{Modeling of Spinning Plates: Geometric Stiffening and Modal Approximation for GNC Applications} }

\author[1, a, *]{ZUCCHELLI Umberto}
\author[1, b, *]{VALLES SÁNCHEZ Irene}
\author[1, c]{SANFEDINO Francesco}
{\tiny \affil[1]{Fédération ENAC ISAE-SUPAERO ONERA, Université de Toulouse, 10 Av. Marc Pélegrin, 31055, Toulouse, France}
\affil[ ]{umberto.zucchelli2@isae-supaero.fr, irene.valles-sanchez@student.isae-supaero.fr, francesco.sanfedino@isae-supaero.fr} }}

\date{}

\begin{document}

\maketitle

% Keywords
\\
{\rmfamily  % Times New Roman 

\noindent \textbf{Abstract} 
\\
This work presents a modal formulation for flexible rectangular plates, accounting for nonlinear geometric effects arising from in-plane foreshortening and centrifugal stiffening. The model is linearized with respect to elastic deformations while retaining the full dependence on spacecraft angular velocities and accelerations. System matrices depend nonlinearly on spacecraft states through squared and cross-product terms, capturing gyroscopic coupling and dynamic stiffening phenomena for arbitrary rotational maneuvers. Polynomial approximation of mode shapes enables efficient computation while preserving accuracy. Model predictions are validated against finite element simulations and literature data for transient response under prescribed hub motion.
\\
 \textbf{Introduction}\\
Lightweight, large-scale flexible structures such as solar arrays and antennas are essential for next-generation space missions. However, their inherent flexibility poses significant challenges for satellite design and control \cite{Liu2007Nonlinear}. Agile manoeuvres can introduce large deformations and geometric stiffening, altering the effective structural stiffness and natural frequencies, potentially destabilizing spacecraft attitude control \cite{Rodrigues2024Modeling, Jinyang2004Geometric, Hoskoti2023Modeling}. Conventional linear models such as Euler--Bernoulli beams or Kirchhoff--Love plates perform well for small deformations, but may underestimate structural stiffness or even exhibit non-physical negative stiffness at high rotational rates \cite{Rodrigues2024Modeling, Hoskoti2023Modeling}. Finite Element Method (FEM) approaches provide high-fidelity predictions but lead to models with a high number of degrees of freedom and computational costs unsuitable for design and control applications. The Assumed Modes Method (AMM) reduces complexity by rstructural deformation as a finite combination of spatial mode shapes and time-dependent generalized coordinates. Previous AMM-based plate formulations, such as those by Yoo et al. \cite{Yoo2001} and Liu and Hong \cite{Jinyang2004}, successfully incorporated geometric stiffening derived from virtual work principles, but omit or simplify the explicit gyroscopic damping terms, limiting accuracy for arbitrary rotational maneuvers.

Recent work by Rodrigues et al. \cite{Rodrigues2024Modeling} developed a Linear Parameter Varying (LPV) formulation for flexible beams that captures state-dependent geometric stiffening and gyroscopic terms across varying spin rates. However, extension to plates introduces additional complexity through in-plane foreshortening and bidirectional bending coupling.
% \cite{alazard_two-input_2015, perez_flexible_2016, sanfedino_finite_2018}

This paper presents a modal formulation for rectangular flexible plates that accounts for nonlinear geometric effects including gyroscopic coupling, centrifugal stiffening, and foreshortening. The derivation follows a Lagrangian approach with quasi-coordinates, linearizing with respect to elastic deformations while retaining the full nonlinear dependence on arbitrary spacecraft angular velocities and accelerations. Polynomial approximation of mode shapes enables efficient symbolic computation of the state-dependent system matrices. %The modal equations are valid for arbitrary rotational maneuvers and provide a foundation for future development of TITOP-based port models suitable for robust control design and modular spacecraft assembly. 
The following section details the modal equations of motion, and mode shape approximation. The results section validates the model against FEM predictions and literature data.

\noindent \textbf{Methodology}\\
\indent \textit{Modal Equation of Motion.}
The dynamics of a flexible solar panel attached to a spinning spacecraft hub are derived using Lagrangian mechanics with quasi-coordinates. The panel is modelled as a thin rectangular plate of length $L$, width $D$, thickness $h$, density $\rho$, Young's modulus $E$, and Poisson's ratio $\nu$, clamped at its root ($\xi=0$) to the hub. The local panel frame $\mathcal{P} = (S; \mathbf{b}_1, \mathbf{b}_2, \mathbf{b}_3)$ has origin at the panel root $S$, with $\xi \in [0,L]$ and $\zeta \in [0,D]$ denoting material coordinates along the length ($\mathbf{b}_1$) and width ($\mathbf{b}_2$) directions. The elastic displacement vector is $\boldsymbol{d}(\xi,\zeta,t) = [d_1, d_2, d_3]^T$, where $d_1, d_2$ are in-plane displacements and $d_3$ is the transverse deflection along $\mathbf{b}_3$.

\noindent Following von Kármán plate theory, geometric nonlinearity arises from foreshortening: when the plate bends ($d_3 \neq 0$), arc length exceeds chord length, inducing in-plane contractions~\cite{Feng2020Large}:
\begin{equation}
d_1(\xi,\zeta,t) = - \int_0^\xi \frac{1}{2}\left(\frac{\partial d_3}{\partial \sigma}\right)^2 d\sigma, \quad d_2(\xi,\zeta,t) = - \int_0^\zeta \frac{1}{2}\left(\frac{\partial d_3}{\partial \tau}\right)^2 d\tau. \label{eq:foreshorten}
\end{equation}

\noindent The transverse deflection is approximated via the assumed modes method with $n$ retained modes:
\begin{equation}
d_3(\xi,\zeta,t) = \sum_{i=1}^{n} q_i(t) W_i(\xi,\zeta) = \boldsymbol{W}(\xi,\zeta)^T \boldsymbol{q}(t), \label{eq:modal_expansion}
\end{equation}
where $\boldsymbol{q}(t) = [q_1, \ldots, q_n]^T \in \mathbb{R}^n$ are time-dependent modal coordinates and $W_i(\xi,\zeta)$ are mode shape functions describing the allowable deformations. Substituting Eq.~\eqref{eq:modal_expansion} into Eq.~\eqref{eq:foreshorten} yields quadratic in-plane displacements $d_1 = - \frac{1}{2}\boldsymbol{q}^T \boldsymbol{H}_1(\xi,\zeta) \boldsymbol{q}$ and $d_2 = - \frac{1}{2}\boldsymbol{q}^T \boldsymbol{H}_2(\xi,\zeta) \boldsymbol{q}$, where $\boldsymbol{H}_1, \boldsymbol{H}_2 \in \mathbb{R}^{n \times n}$ are symmetric foreshortening matrices with elements:
\begin{equation}
H_{1,ij}(\xi,\zeta) = \int_0^\xi \frac{\partial W_i}{\partial \sigma} \frac{\partial W_j}{\partial \sigma} d\sigma, \quad H_{2,ij}(\xi,\zeta) = \int_0^\zeta \frac{\partial W_i}{\partial \tau} \frac{\partial W_j}{\partial \tau} d\tau. \label{eq:H_def}
\end{equation}

\noindent Denote the position vector from hub center $B$ to material point $Q$ at coordinates $(\xi,\zeta)$ in panel frame $\mathcal{P}$ as $\boldsymbol{r}^{BQ}_\mathcal{P} = [A_l+\xi+d_1, B_l+\zeta-D/2+d_2, C_l+d_3]^T$, where $[A_l, B_l, C_l]^T$ is the lever arm from $B$ to panel root $S$ and the term $-D/2$ accounts for the panel being attached at its mid-width point. The velocity of $Q$ in panel frame is $\boldsymbol{v}^Q_\mathcal{P} = \boldsymbol{v}^B_\mathcal{P} + \boldsymbol{\omega}^\mathcal{P} \times \boldsymbol{r}^{BQ}_\mathcal{P} + \dot{\boldsymbol{d}}_\mathcal{P}$, where $\boldsymbol{v}^B_\mathcal{P}$ is hub velocity, $\boldsymbol{\omega}^\mathcal{P} = [\omega_1, \omega_2, \omega_3]^T$ is angular velocity, and $\dot{\boldsymbol{d}}_\mathcal{P} = [\dot{\boldsymbol{q}}^T\boldsymbol{H}_1\boldsymbol{q}, \dot{\boldsymbol{q}}^T\boldsymbol{H}_2\boldsymbol{q}, \boldsymbol{W}^T\dot{\boldsymbol{q}}]^T$ is the material derivative of displacement. The kinetic and potential energies are:
\begin{equation}
\mathcal{T} = \frac{1}{2}\int_0^L \int_0^D \rho h \, \boldsymbol{v}^{Q,T}_\mathcal{P} \boldsymbol{v}^Q_\mathcal{P} \, d\zeta \, d\xi, \quad \mathcal{V} = \frac{1}{2}\int_0^L \int_0^D D_{\text{flex}}\left[(\nabla^2 d_3)^2 - 2(1-\nu)\left(\frac{\partial^2 d_3}{\partial\xi\partial\zeta}\right)^2\right] d\zeta \, d\xi, \label{eq:energies}
\end{equation}
where $\rho h = \rho_\delta$ is area density and $D_{\text{flex}} = Eh^3/[12(1-\nu^2)]$ is flexural rigidity.

\noindent The Lagrangian $\mathcal{L} = \mathcal{T} - \mathcal{V}$ is linearized with respect to modal coordinates $\boldsymbol{q}$ around the undeformed configuration $\overline{\boldsymbol{q}} = \boldsymbol{0}$ using second-order Taylor expansion in quasi-coordinates. 
%Crucially, no linearization is performed with respect to spacecraft angular velocities $\boldsymbol{\omega}^{\mathcal{P}}$ or accelerations $\dot{\boldsymbol{\omega}}^{\mathcal{P}}$, which remain arbitrary time-varying quantities. 
Applying the Lagrange-d'Alembert principle yields the $n$ modal equations of motion:
\begin{equation}
\boldsymbol{M}^{\mathcal{P}} \ddot{\boldsymbol{q}} + 2\boldsymbol{D}_{\text{gyro}}^{\mathcal{P}}(\boldsymbol{\omega}^{\mathcal{P}})\dot{\boldsymbol{q}} + \left[\boldsymbol{K}_{sp}^{\mathcal{P}} + \boldsymbol{K}_d^{\mathcal{P}}(\boldsymbol{\omega}^{\mathcal{P}}, \boldsymbol{\dot\omega}^{\mathcal{P}})\right]\boldsymbol{q}  = \boldsymbol{P}^{\mathcal{P}T}\dot{\boldsymbol{\omega}}^{\mathcal{P}} + \boldsymbol{L}^{\mathcal{P}T}\boldsymbol{a}^{\mathcal{P}} + \boldsymbol{Q}_{\text{ext}}, \label{eq:modal_eom}
\end{equation}
where $\dot{\boldsymbol{\omega}}^{\mathcal{P}} = [\dot\omega_1, \dot\omega_2, \dot\omega_3]^T \in \mathbb{R}^3$ is angular acceleration in panel frame, $\boldsymbol{a}^{\mathcal{P}} \in \mathbb{R}^3$ is hub linear acceleration, and $\boldsymbol{Q}_{\text{ext}} \in \mathbb{R}^n$ represents external generalized forces. Although linear in modal coordinates $\boldsymbol{q}$, the system matrices depend nonlinearly on arbitrary spacecraft angular velocities through squared terms ($\omega_i^2$) and cross-products ($\omega_i \omega_j$), capturing geometric stiffening and gyroscopic effects for any rotational maneuver.

\indent \textit{System Matrices.}
The modal mass matrix $\boldsymbol{M}^{\mathcal{P}} \in \mathbb{R}^{n \times n}$ is:
\begin{equation}
M_{ij}^{\mathcal{P}} = \iint_{A} \rho_\delta W_i(\xi,\zeta) W_j(\xi,\zeta) \, d\zeta \, d\xi. \label{eq:mass_matrix}
\end{equation}

\noindent The gyroscopic damping matrix $\boldsymbol{D}_{\text{gyro}}^{\mathcal{P}}(\boldsymbol{\omega}^{\mathcal{P}}) \in \mathbb{R}^{n \times n}$ represents Coriolis forces, is linear in $\dot{\boldsymbol{q}}$, and must be skew-symmetric ($\boldsymbol{D}_{\text{gyro}}^T = -\boldsymbol{D}_{\text{gyro}} = \frac{1}{2}(\tilde{\boldsymbol{D}} - \tilde{\boldsymbol{D}}^T)$) to ensure energy conservation:
\begin{equation}
\tilde{D}_{ij} = 
\iint_A \rho_\delta \, W_i \Bigg[ \omega_3 \Big( (B_l+\zeta-D/2) \frac{\partial W_j}{\partial \xi} - (A_l+\xi) \frac{\partial W_j}{\partial \zeta} \Big) + C_l \Big( \omega_2 \frac{\partial W_j}{\partial \zeta} - \omega_1 \frac{\partial W_j}{\partial \xi} \Big) \Bigg] \, d\zeta \, d\xi,
\label{eq:D_gyro_raw}
\end{equation}

\noindent The static stiffness matrix $\boldsymbol{K}_{sp}^{\mathcal{P}} \in \mathbb{R}^{n \times n}$ arises from plate bending energy:
\begin{align}
K_{sp,ij}^{\mathcal{P}} = \iint_A D_{\text{flex}}  &\left[\frac{\partial^2 W_i}{\partial \xi^2}\frac{\partial^2 W_j}{\partial \xi^2} + \frac{\partial^2 W_i}{\partial \zeta^2}\frac{\partial^2 W_j}{\partial \zeta^2} + \nu\left(\frac{\partial^2 W_i}{\partial \xi^2}\frac{\partial^2 W_j}{\partial \zeta^2} + \frac{\partial^2 W_i}{\partial \zeta^2}\frac{\partial^2 W_j}{\partial \xi^2}\right) \right. \nonumber \\
&\left. + 2(1-\nu)\frac{\partial^2 W_i}{\partial \xi \partial \zeta}\frac{\partial^2 W_j}{\partial \xi \partial \zeta}\right] d\zeta \, d\xi. \label{eq:static_stiffness}
\end{align}

\noindent The dynamic stiffness matrix $\boldsymbol{K}_d^{\mathcal{P}}(\boldsymbol{\omega}^{\mathcal{P}}, \dot{\boldsymbol{\omega}}^{\mathcal{P}}) \in \mathbb{R}^{n \times n}$ captures geometric stiffening effects from centrifugal and Euler forces:
\begin{equation}
K_{d,ij}^{\mathcal{P}} = \iint_A \rho_\delta I_{ij}(\xi, \zeta, t) \, d\zeta \, d\xi, \label{eq:dynamic_stiffness}
\end{equation}
where the integrand contains centrifugal (quadratic in $\boldsymbol{\omega}$), Euler (linear in $\dot{\boldsymbol{\omega}}$), and cross-product terms:
\begin{align}
I_{ij} = &\Big\{C_l(\dot{\omega}_2+\omega_1\omega_3) - [B_l+\zeta-D/2](\dot{\omega}_3-\omega_1\omega_2) - [A_l+\xi](\omega_2^2+\omega_3^2)\Big\} H_{1,ij} \nonumber \\
&+ \Big\{[A_l+\xi](\dot{\omega}_3+\omega_1\omega_2) - C_l(\dot{\omega}_1-\omega_3\omega_2) - [B_l+\zeta-D/2](\omega_1^2+\omega_3^2)\Big\} H_{2,ij} \nonumber \\
&- (\omega_1^2+\omega_2^2) W_i W_j. \label{eq:I_ij_integrand}
\end{align}

\noindent The inertial coupling matrices $\boldsymbol{P}^{\mathcal{P}}, \boldsymbol{L}^{\mathcal{P}} \in \mathbb{R}^{3 \times n}$ couple hub accelerations to modal dynamics. Matrix $\boldsymbol{P}^{\mathcal{P}}$ arises from angular acceleration $\dot{\boldsymbol{\omega}}^\mathcal{P}$ acting through lever arm and position, while $\boldsymbol{L}^{\mathcal{P}}$ couples linear acceleration $\boldsymbol{a}^\mathcal{P}$ to flexible modes:
\begin{equation}
P_{1i}^{\mathcal{P}} = \iint_A \rho_\delta (B_l+\zeta - D/2) W_i \, dA, \quad P_{2i}^{\mathcal{P}} = - \iint_A \rho_\delta (A_l+\xi) W_i \, dA, \quad P_{3i}^{\mathcal{P}} = 0, \label{eq:P_explicit}
\end{equation}
\begin{equation}
L_{ji}^{\mathcal{P}} = \rho h \iint_A W_i \, dA \cdot \hat{\mathbf{b}}_j, \label{eq:L_explicit}
\end{equation}
where $\hat{\mathbf{b}}_j$ denotes the unit vector along direction $j$ in frame $\mathcal{P}$.

\indent \textit{Mode Shapes and Polynomial Approximation.}
Mode shapes are constructed as separable products $W_i(\xi,\zeta) = \Phi_{m(i)}(\xi)\Psi_{n(i)}(\zeta)$, combining four clamped-free (CF) $\Phi$ and four free-free (FF) $\Psi$ Euler-Bernoulli beam modes in each direction, where the first two FF modes represent rigid-body translation and rotation, yielding $n = 16$ modes. Since analytical modes involve transcendental functions, each mode is approximated by constrained least-squares polynomial fits enforcing boundary conditions and achieving RMSE $<10^{-5}$. This reduces symbolic integration time from hours to seconds. Figures~\ref{fig:fitted_modes_1D} and~\ref{fig:combined_modes_3D} show the polynomial-fitted modes and resulting 3D combined shapes.

\begin{figure}[h!]
\centering
\begin{minipage}[t]{0.48\textwidth}
\centering
\includegraphics[width=\textwidth]{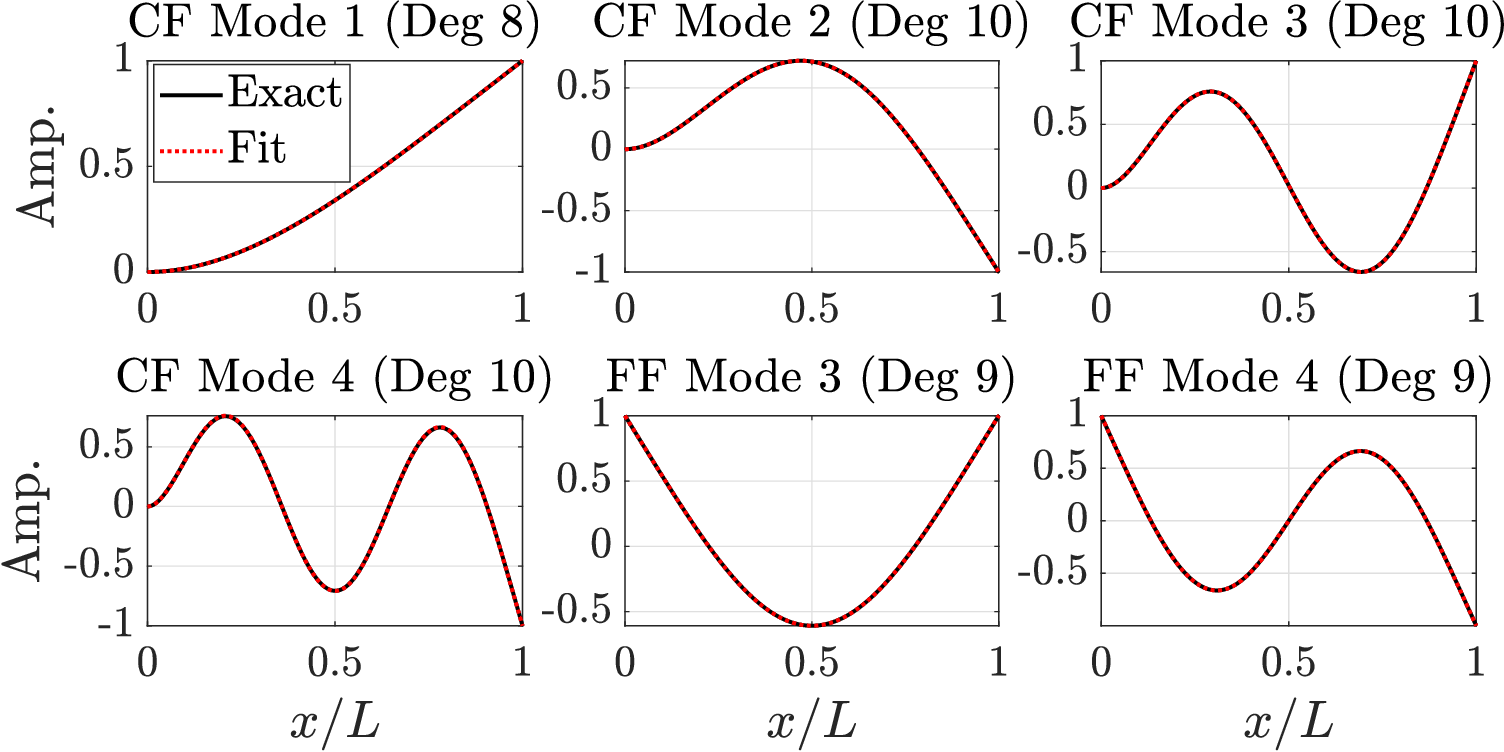}
\caption{\textit{Example of polynomial-fitted CF modes $\Phi_m(\xi)$ and FF modes $\Psi_n(\zeta)$ compared to exact analytical solutions.}}
\label{fig:fitted_modes_1D}
\end{minipage}
\hfill
\begin{minipage}[t]{0.48\textwidth}
\centering
\includegraphics[width=\textwidth]{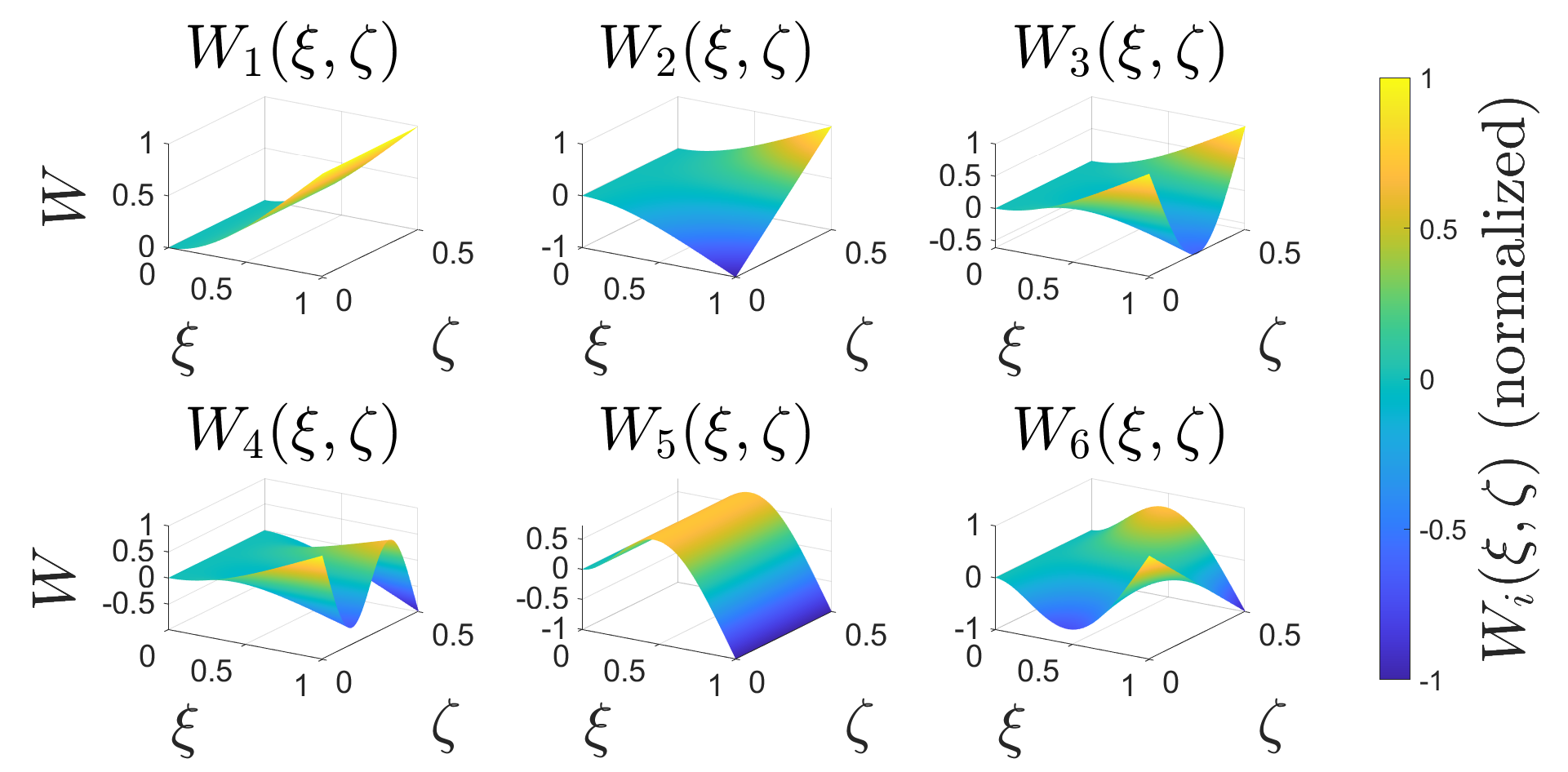}
\caption{\textit{Examples of 3D combined mode shapes $W_i(\xi,\zeta) = \Phi_{m(i)}(\xi)\Psi_{n(i)}(\zeta)$.}}
\label{fig:combined_modes_3D}
\end{minipage}
\end{figure}

\noindent {\textbf{Results}}\\
The present section validates the proposed modal formulation by comparing its predictions for the evolution of natural frequencies with spinning rate against FEM results, and the transient tip displacement responses under prescribed hub motion against data from the literature.

\indent \textit{Natural Frequency Evolution.}
The natural frequencies of the spinning plate are obtained by solving the generalized eigenvalue problem:
\begin{equation}
\det\left(\boldsymbol{K}_{sp}^{\mathcal{P}} + \Omega^2 \boldsymbol{K}_{d_0}^{\mathcal{P}} - \omega^2 \boldsymbol{M}^{\mathcal{P}}\right) = 0, \label{eq:eigenvalue_problem}
\end{equation}
where $\omega$ represents the natural frequency at spinning rate $\Omega$ about the panel root axis $\mathbf{b}_2$. The normalized dynamic stiffness matrix $\boldsymbol{K}_{d_0}^{\mathcal{P}}$ is evaluated at the equilibrium spinning state $\overline{\boldsymbol{\omega}}^{\mathcal{P}} = [0, \Omega, 0]^T$ such that $\boldsymbol{K}_d^{\mathcal{P}}(\overline{\boldsymbol{\omega}}^{\mathcal{P}}, \boldsymbol{0}) = \Omega^2 \boldsymbol{K}_{d_0}^{\mathcal{P}}$. 

\noindent The plate parameters used for this analysis are: length $L = 1.0$ m, width $D = 0.5$ m, thickness $h = 0.0025$ m, density $\rho = 3000.0$ kg/m$^3$, Young's modulus $E = 70.0$ GPa, and Poisson's ratio $\nu = 0.3$. Figure~\ref{fig:frequency_evolution} illustrates the evolution of the first four natural frequencies as functions of the spinning rate $\Omega$. The frequencies exhibit a quadratic increase, consistent with the $\Omega^2$ scaling in the dynamic stiffness matrix $\boldsymbol{K}_d^{\mathcal{P}}$. The results show good agreement with NASTRAN finite element predictions and demonstrate that the polynomial mode shape approximation preserves the fundamental physics of spinning structures.

% \begin{figure}[H]
% \centering
% \includegraphics[width=0.9\textwidth]{Images/4_freq_FEM.eps}
% \caption{\textit{Evolution of the first four natural frequencies as functions of spinning rate $\Omega \in [0, 50]$ rad/s.}}
% \label{fig:frequency_evolution}
% \end{figure}

\indent \textit{Transient Simulation.}
To validate the transient response predictions, the model is compared against results from Yoo et al. \cite{Yoo2001}, and  Liu and Hong \cite{Jinyang2004}, who derived flexible plate equations using a virtual work principle. Their formulation includes geometric nonlinearities and dynamic stiffening but does not explicitly account for gyroscopic damping terms. 

\noindent The prescribed hub motion from \cite{Yoo2001, Jinyang2004} is a smooth angular acceleration about the panel root axis $\mathbf{b}_2$:
\begin{equation}
\omega_1(t) = \omega_3(t) = 0, \quad \omega_2(t) =
    \begin{cases}
    \omega_0^* + \omega_s^* \left( \dfrac{t}{t_s^*} - \dfrac{1}{2\pi} \sin\left( \dfrac{2\pi t}{t_s^*} \right) \right), & t \leq t_s^* \\
    \omega_0^* + \omega_s^*, & t > t_s^*
    \end{cases} \label{eq:prescribed_motion}
\end{equation}
where $\omega_0^* = 0$ rad/s is the initial angular velocity, $\omega_s^* = 10$ rad/s is the amplitude of the angular velocity ramp-up, and $t_s^* = 5$ s is the transition time. The panel parameters are identical to those used in the frequency evolution analysis and presented in \cite{Yoo2001, Jinyang2004}. The modal equations~\eqref{eq:modal_eom} are integrated numerically using a fourth-order Runge-Kutta scheme with time step $\Delta t = 10^{-4}$ s.

Figure~\ref{fig:transient_response} presents the transient tip displacement at position $(\xi, \zeta) = (L, D/2)$ over a 10-second interval. The present model reproduces the main dynamic features reported in the literature, exhibiting a maximum deflection around 2.5 s. After approximately 5 s, the response oscillates about the equilibrium position, exhibiting characteristics consistent with the literature results.

% \begin{figure}[H]
% \centering
% \includegraphics[width=0.7\textwidth]{Images/lat_displ_lit.eps}
% \caption{\textit{Transient tip displacement at $(\xi, \zeta) = (L, D/2)$ under prescribed hub motion. Solid line: present model; dashed line: Liu et al. \cite{Jinyang2004}, and Yoo et al. \cite{Yoo2001}.}}
% \label{fig:transient_response}
% \end{figure}

\begin{figure}[h!]
\centering
\begin{minipage}[t]{0.48\textwidth}
\centering
\includegraphics[width=\textwidth]{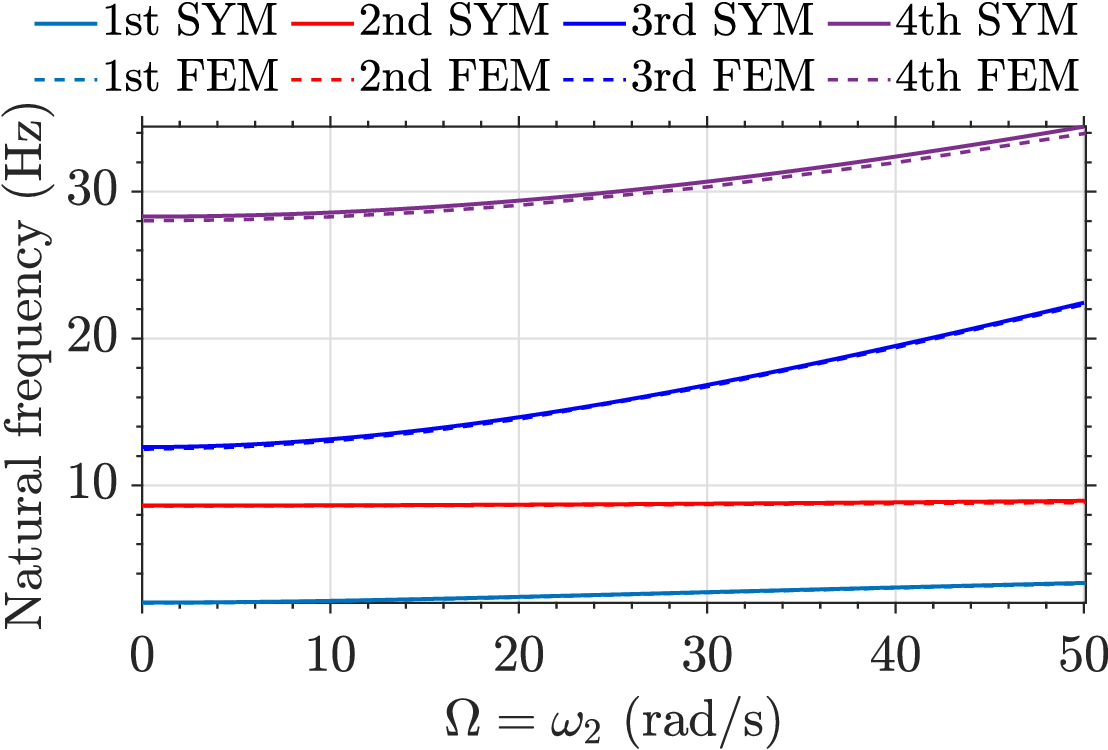}
\caption{\textit{Evolution of the first four natural frequencies as functions of spinning rate $\Omega \in [0, 50]$ rad/s.}}
\label{fig:frequency_evolution}
\end{minipage}
\hfill
\begin{minipage}[t]{0.48\textwidth}
\centering
\includegraphics[width=\textwidth]{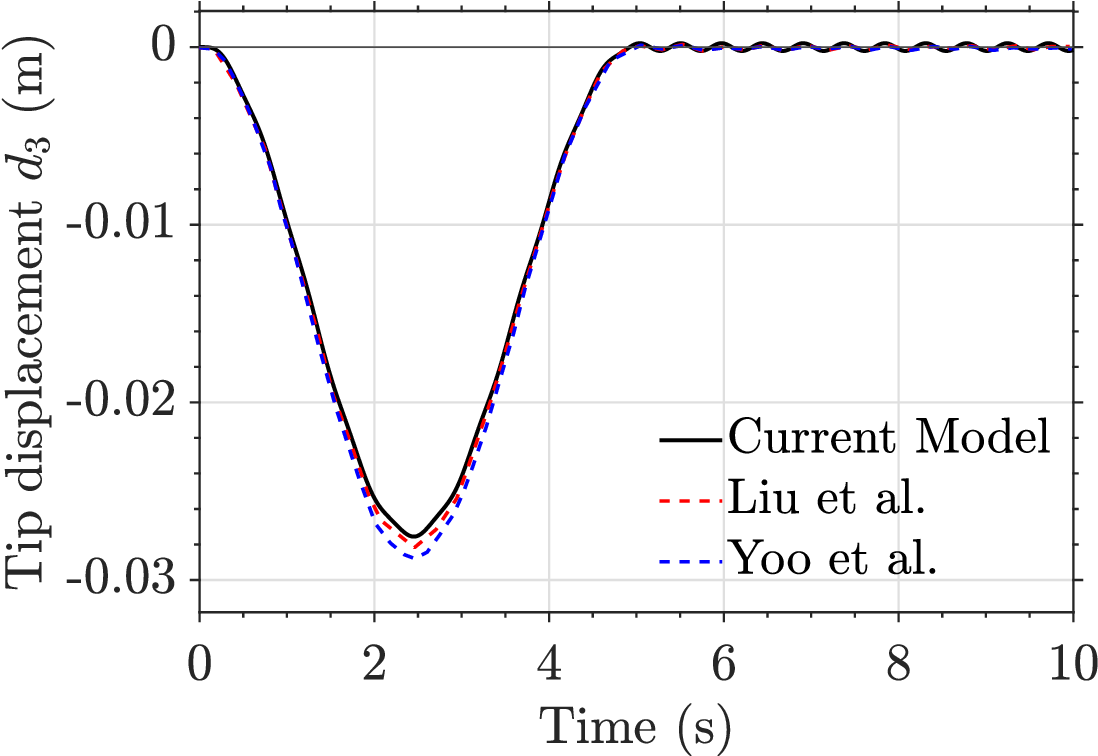}
\caption{\textit{Transverse tip displacement $d_3(L, D/2, t)$ normal to the panel, under prescribed hub motion \eqref{eq:prescribed_motion}.}}
\label{fig:transient_response}
\end{minipage}
\end{figure}

\noindent {\textbf{Conclusion}}\\
This work presented a modal formulation for spinning flexible rectangular plates that accounts for nonlinear geometric effects including gyroscopic coupling, centrifugal stiffening, and in-plane foreshortening. The Lagrangian approach with quasi-coordinates yields modal equations linear in modal coordinates but with system matrices that depend nonlinearly on spacecraft angular velocities and accelerations through squared and cross-product terms. A polynomial approximation strategy for mode shapes enables efficient closed-form computation of these state-dependent matrices while maintaining high accuracy, reducing symbolic integration time. Validation against FEM simulations demonstrates accurate prediction of natural frequency evolution with spinning rate. Comparison with literature data for transient response under prescribed hub motion confirms the model's ability to reproduce key dynamic features. The modal equations~\eqref{eq:modal_eom} provide a foundation for future development of control-oriented frameworks such as the Two-Input Two-Output Port (TITOP) approach \cite{alazard_two-input_2015, perez_flexible_2016, sanfedino_finite_2018}, enabling LPV control synthesis and modular spacecraft assembly with flexible appendages.

\bibliographystyle{unsrt}
\bibliography{references}

@article{Liu2007Nonlinear,
  author = {Liu, J. and Hong, J.},
  title = {Nonlinear Formulation for Flexible Multibody System with Large Deformation},
  journal = {Acta Mech. Sin.},
  volume = {23},
  pages = {111--119},
  year = {2007}
}

@article{Feng2020Large,
  author = {Feng, L. and Baozeng, Y. and Banerjee, A. K.},
  title = {Large Motion Dynamics of In-Orbit Flexible Spacecraft with Large-Amplitude Propellant Slosh},
  journal = {J. Guid. Control Dyn.},
  volume = {43},
  number = {3},
  year = {2020}
}

@article{Rodrigues2024Modeling,
  author = {Rodrigues, R. and Alazard, D. and Sanfedino, F. and Mauriello, T. and Iannelli, P.},
  title = {Modeling and Analysis of a Flexible Spinning Euler-Bernoulli Beam with Centrifugal Stiffening and Softening: A Linear Fractional Representation Approach with Application to Spinning Spacecraft},
  journal = {Appl. Math. Model.},
  year = {2024}
}

@article{Jinyang2004,
  author  = {Liu, J. and Hong, J.},
  title   = {Geometric Nonlinear Formulation and Discretization for a Rectangular Plate Undergoing Overall Motions},
  journal = {Mech. Res. Commun.},
  volume  = {32},
  pages   = {561--571},
  year    = {2004}
}

@article{Hoskoti2023Modeling,
  author = {Hoskoti, L. and Gupta, S. S. and Sucheendran, M. M.},
  title = {Modeling of Geometrical Stiffening in a Rotating Blade—A Review},
  journal = {J. Sound Vib.},
  volume = {548},
  pages = {117526},
  year = {2023}
}

@article{Yoo2001,
  author  = {Yoo, H. H. and Chung, J.},
  title   = {Dynamics of Rectangular Plates Undergoing Prescribed Overall Motion},
  journal = {J. Sound Vib.},
  volume  = {239},
  number  = {1},
  pages   = {123--137},
  year    = {2001}
}

@inproceedings{alazard_two-input_2015,
	address = {Kissimmee, Florida},
	title = {Two-Input Two-Output Port Model for Mechanical Systems},
	doi = {10.2514/6.2015-1778},
	booktitle = {AIAA Guidance, Navigation, and Control Conference},
	publisher = {American Institute of Aeronautics and Astronautics},
	author = {Alazard, D. and Perez, J. A. and Cumer, C. and Loquen, T.},
	year = {2015}
}

@article{perez_flexible_2016,
	title = {Flexible Multibody System Linear Modeling for Control Using Component Modes Synthesis and Double-Port Approach},
	volume = {138},
	doi = {10.1115/1.4034149},
	journal = {J. Dyn. Syst. Meas. Control},
	author = {Perez, J. A. and Alazard, D. and Loquen, T. and Pittet, C. and Cumer, C.},
	year = {2016}
}

@article{sanfedino_finite_2018,
	title = {Finite Element Based N-Port Model for Preliminary Design of Multibody Systems},
	volume = {415},
	doi = {10.1016/j.jsv.2017.11.021},
	journal = {J. Sound Vib.},
	author = {Sanfedino, F. and Alazard, D. and Pommier-Budinger, V. and Falcoz, A. and Boquet, F.},
	year = {2018},
	pages = {128--146}
}

@article{Jinyang2004Geometric,
  author = {Jinyang, L. and Jiazhen, H.},
  title = {Geometric nonlinear formulation and discretization for a rectangular plate undergoing overall motions},
  journal = {Mechanics Research Communications},
  number = {32},
  pages = {561--571},
  year = {2004}
}

\end{document}